\documentclass[preprint2]{aastex}

\shorttitle{X-ray signature from coronal electron beams}
\shortauthors{Saint-Hilaire et al.}

\begin{document}

\title{X-ray emission from the base of a current sheet in the wake of a CME}

\author{P. Saint-Hilaire\altaffilmark{1}, S. Krucker\altaffilmark{1}, and R.P. Lin\altaffilmark{1}}
\affil{Space Sciences Laboratory, University of California,
    Berkeley, CA 94720}

\email{shilaire@ssl.berkeley.edu}

\begin{abstract}
	Following a CME which started on 2002 November 26, RHESSI, the {\it Ramaty High Energy Solar Spectroscopic Imager}, 
	observed for 12 hours an X-ray source above the solar limb, at altitudes between 0.1 and 0.3 $R_S$ above the photosphere.
	The GOES baseline was remarkably high throughout this event.
	The X-ray source's temperature peaked around 10--11 MK, and its emission measure increased throughout this time interval.
	Higher up, at 0.7 $R_S$, hot (initially $>$8 MK) plasma has been observed by UVCS on SoHO for 2.3 days.
	This hot plasma was interpreted as the signature of a current sheet trailing the CME \citep{Bemporad2006}.

	The thermal energy content of the X-ray source is more than an order of magnitude larger than in the current sheet.
	Hence, it could be the source of the hot plasma in the current sheet, 
	although current sheet heating by magnetic reconnection within it cannot be discounted.

	To better characterize the X-ray spectrum, we have used novel techniques (back-projection-based and visibility-based) for long integration (several hours) imaging spectroscopy.
	There is no observed non-thermal hard X-ray bremsstrahlung emission, leading to the conclusion that there is either very little particle acceleration occurring in the vicinity of this post-flare X-ray source,
	or that either the photon spectral index would have had to be uncharacteristically (in flare parlance) high ($\gamma \gtrsim$8) and/or the low-energy cutoff very low ($E_c \lesssim$6 keV).

\end{abstract}


%+++++++++++++++++++++++++++++++++++++++++++++++++++++++++++++++++++++++++++++++++++++++++++++++++++++++++++++++++++++
\section{Introduction}

During solar flares, particles are believed to be accelerated, and plasma heated as a result of magnetic reconnection at an X-point or neutral sheet in the corona \citep{Kopp1976}.
The accelerated electrons stream down to the footpoints of coronal magnetic loops, producing hard X-ray bremsstrahlung as they are thermalized by Coulomb collisions
in the dense lower corona or chromosphere.
The directly heated plasma already in the loop and the ablated chromospheric material produce hot loops below the reconnection site, with temperature that can be 20 MK or higher, 
and densities as high as 10$^{11}$ cm$^{-3}$.
These loops are visible in soft X-rays, and later, as they cool down, become visible in EUV and in H$\alpha$.
The reconnection site gradually moves upwards and continues to release energy, even as the X-ray flux diminishes.
This translates into the appearance of higher and higher hot loops, and cooler loops at the lower altitudes \citep[see e.g.][]{Svetska1987}.

Coronal mass ejections (CMEs) are often associated with flares.
One of the models invoked in their creation is the catastrophe or flux-rope model \citep[see e.g.][]{LinForbes2000},
in which a current sheet (CS) is thought to extend from the top of the reconnected loop system to the plasma bubble that surrounds the expelled flux rope.
A CS is supposed to be so thin as to make direct observation quite difficult.
However, there have recently been reports of CS detection in the extended corona from observations acquired in the wake of CMEs by the 
{\it Ultraviolet Coronograph Spectrometer} \citep[UVCS;][]{Kohl1995}, in the form of narrow, very hot (several MK) features, most prominently in the Fe$^{17+}$ line: 
\citet{Ciaravella2002,Ko2003,Raymond2003,Lin2005, Bemporad2006, Bemporad2008, Ciaravella2008}.
In particular, \citet{Ciaravella2008} have firmly established that the CS thickness (for one event, at least) to be between 0.04 and 0.08 $R_S$, 
far larger than classical ($\lesssim$100 m) or anomalous (a few 10s' of km) resistivity would predict.
In a re-analysis of previous results, \citet{Bemporad2008} have explained these observations with the existence of many 
($\sim$10$^{-11}$ to 10$^{-17}$ m$^{-3}$) microscopic CSs of small sizes ($\approx$10--10$^4$ m) that, through non-thermal turbulent broadening, 
can justify not only the high CS temperatures but also the large observed thicknesses of macroscopic CSs.
 	
\citet{Bemporad2006} examine one such event, that lasted at least 2.3 days. 
The CME to which these observations pertain started at around 17:00 UT on 2002 November 26 on the western limb of the Sun.
\citet{Bemporad2006} discusses in detail UVCS, {\it Large Angle and Spectrometric Coronograph} \citep[LASCO;][]{Brueckner1995} and {\it Extreme Ultraviolet Imager} \citep[EIT;][]{Delaboudiniere1995} (instruments on board SOHO, the {\it Solar and Heliospheric Observatory}) observations of this event.
Our paper will concentrate on examining the concurrent X-ray emission, with data from {\it GOES} (Geostationary Environmental Satellite), and the {\it Ramaty High Energy Solar Spectroscopic Imager} \citep[RHESSI][]{Lin2002},
which was launched a few months prior to this event.

Section~\ref{sect:obs} will briefly summarize the observations reported by \citet{Bemporad2006}, then complement them with X-rays observations: lightcurves, spectra and imaging.
Section~\ref{sect:discussion} will then discuss interpretations of these observations, and the possibility that accelerated non-thermal particles provide the energy required to power the X-ray source, and perhaps the CS.

%+++++++++++++++++++++++++++++++++++++++++++++++++++++++++++++++++++++++++++++++++++++++++++++++++++++++++++++++++++++
\section{Observations} \label{sect:obs}

\subsection{Brief summary of previously reported observations:}

	\citet{Bemporad2006} have reported observing a current sheet (CS) in the wake of a CME that started on 2002 November 26 around 17:00 UT.
	That conclusion was mainly supported by UVCS observations (starting at 18:39 UT) of a hot (initially well beyond 8 MK) plasma
	above the western limb of the Sun ($\approx 25^{\circ}$ north latitude), in the same radial direction as the CME, 
	at an altitude of about 0.7 $R_S$ above the solar photosphere, directly above a loop system observed with EIT.
	This hot plasma had a width of $\approx$100 Mm perpendicularly to the radial direction from the Sun and to our line of sight.
	It cooled to 3.5 MK after 2.3 days, at which point UVCS observations stopped.
	
	\citet{Bemporad2006} also estimated that adiabatic heating is insufficient to explain the hot plasma, at least initially, and that reconnection must be the source of the thermal energy.
	In his re-analysis, \citet{Bemporad2008} further strengthens that hypothesis.
		%attributed the decrease of CS heating to the decrease in intensity of the (presumably reconnecting) ambient magnetic field.
		
	The remainder of this section will concentrate on complementing the aforementioned study with X-ray observations from RHESSI and GOES.

\subsection{X-ray lightcurves and imaging}

		\begin{figure*}[ht!]
		\centering
		\includegraphics[width=16.6cm]{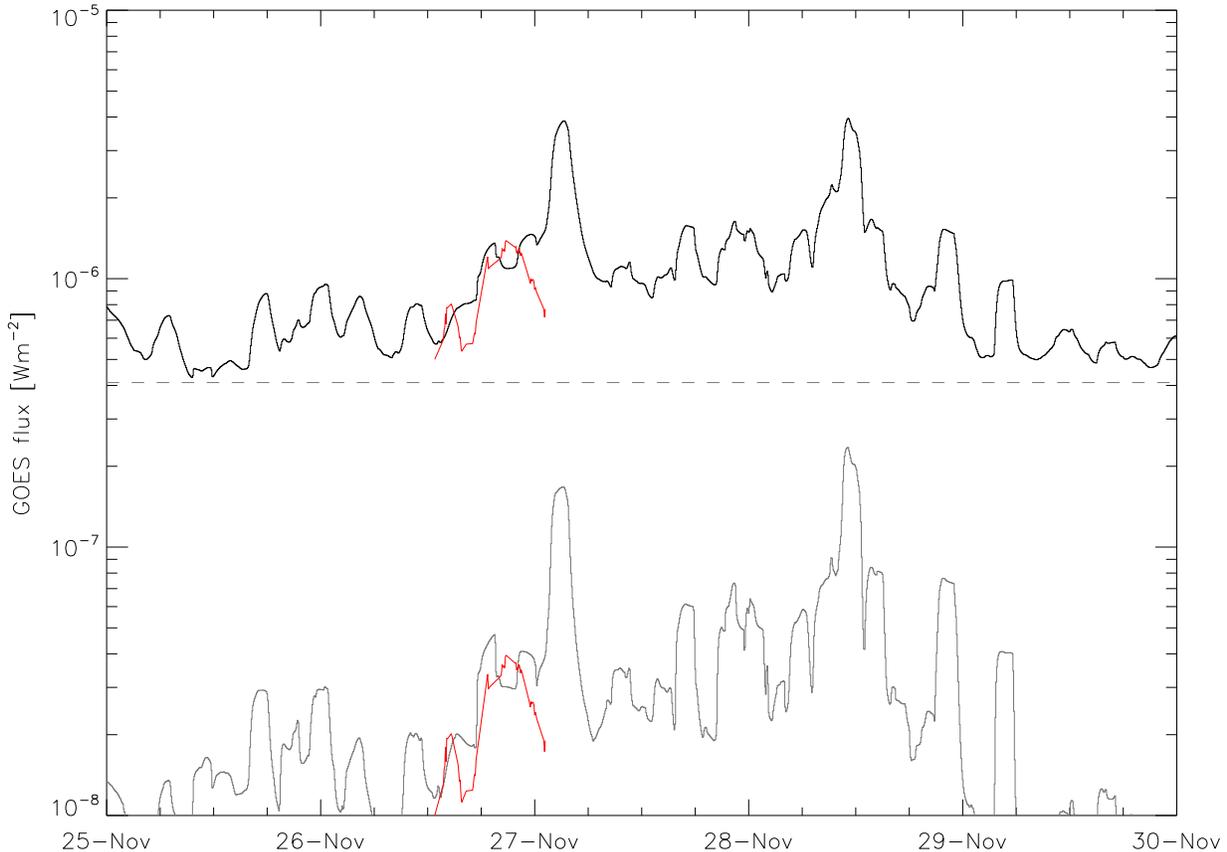}
		\caption{
			GOES lightcurves: {\it black} is the 1--8$\AA$ flux, the {\it gray} line is the 0.5--4$\AA$ flux.
			The data has been smoothed using a 2-hour smoothing window.
			The {\it dashed line} represents a constant flux at 4.1$\times$10$^{-7}$ W m$^{-2}$.
			The {\it red line} represents RHESSI 4--8 keV flux {\it from the coronal source only}.
		}
		\label{fig:goes}
		\end{figure*}

\begin{figure*}[ht!]
		\centering
		\includegraphics[height=15cm]{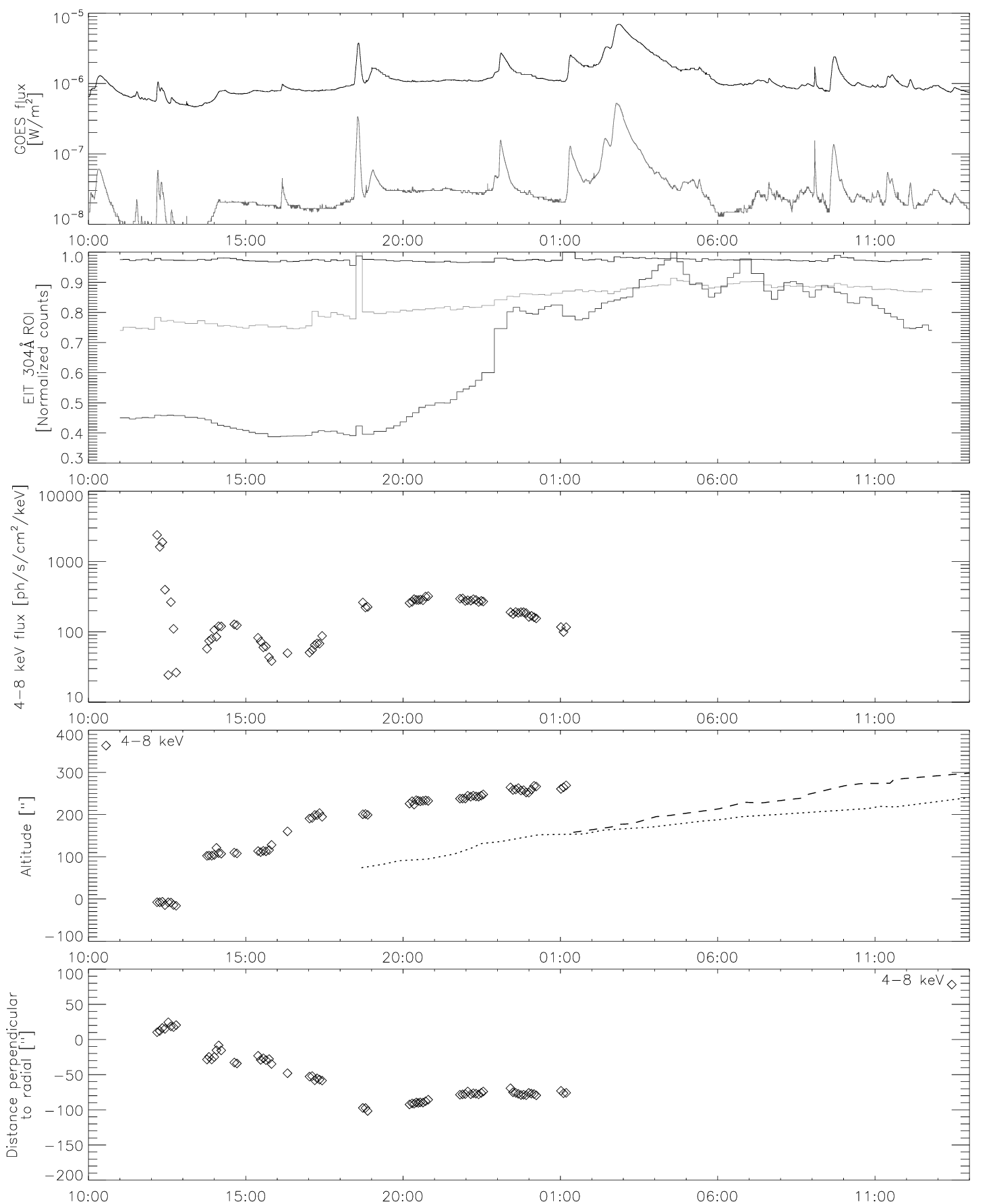}
		\caption{
			{\it First (top) plot:} GOES lightcurves ({\it black:} 1--8 {\AA}, {\it gray:} 0.5--4 {\AA}).
			{\it Second plot:} EIT lightcurves ({\it black:} full-Sun, {\it light gray:} ROI is 600'' square centered around [1050,450], i.e. encompassing slightly more than Figure~\ref{fig:compositeimg}, 
				{\it dark gray:} ROI is 200''x300'' rectangle centered at [1050,450]), shown as a green box in Figure~\ref{fig:compositeimg}.
			{\it Third plot:} RHESSI 4--8 keV flux from imaging with ROI being a 256'' square centered around [1100,400].
			{\it Fourth plot:} Diamonds: source altitude in the 4--8 keV band, from RHESSI imaging using subcollimator 8.
				Dashed line: northern $EIT$ loop system altitude \citep[from][]{Bemporad2006}.
				Dotted line: southern $EIT$ loop system altitude \citep[from][]{Bemporad2006}.
			{\it Fifth plot:} Source azimuthal distance from an arbitrary radial, using RHESSI's subcollimator 8 in the 4--8 keV band.
			%Spread around 13:00 is due to rear antenna noise. After 2002/11/27  01:10: source is ``lost''.
			In the RHESSI plots (last three plots), the displayed information include the initial flare at $\approx$12:00 UT.
			Beyond  13:00 UT, the plots only display the information pertaining to the coronal source, removing any disc flares.
		}
		\label{fig:tp}
		\end{figure*}

		As can be seen in Figure~\ref{fig:goes} and~\ref{fig:tp}, on 2002 November 26, around 13:40 UT, 
		the GOES ``baseline'' in both channels increased suddenly, and stayed fairly high until 2002 November 29 $\approx$00:00 UT.
		During this time interval, several flares occurred at different positions on the solar disc, 
		revealed as individual peaks in the GOES lightcurves (which are spatially integrated) of Figure~\ref{fig:tp}.

		\begin{figure}[ht!]
		\centering
		\includegraphics[width=7cm]{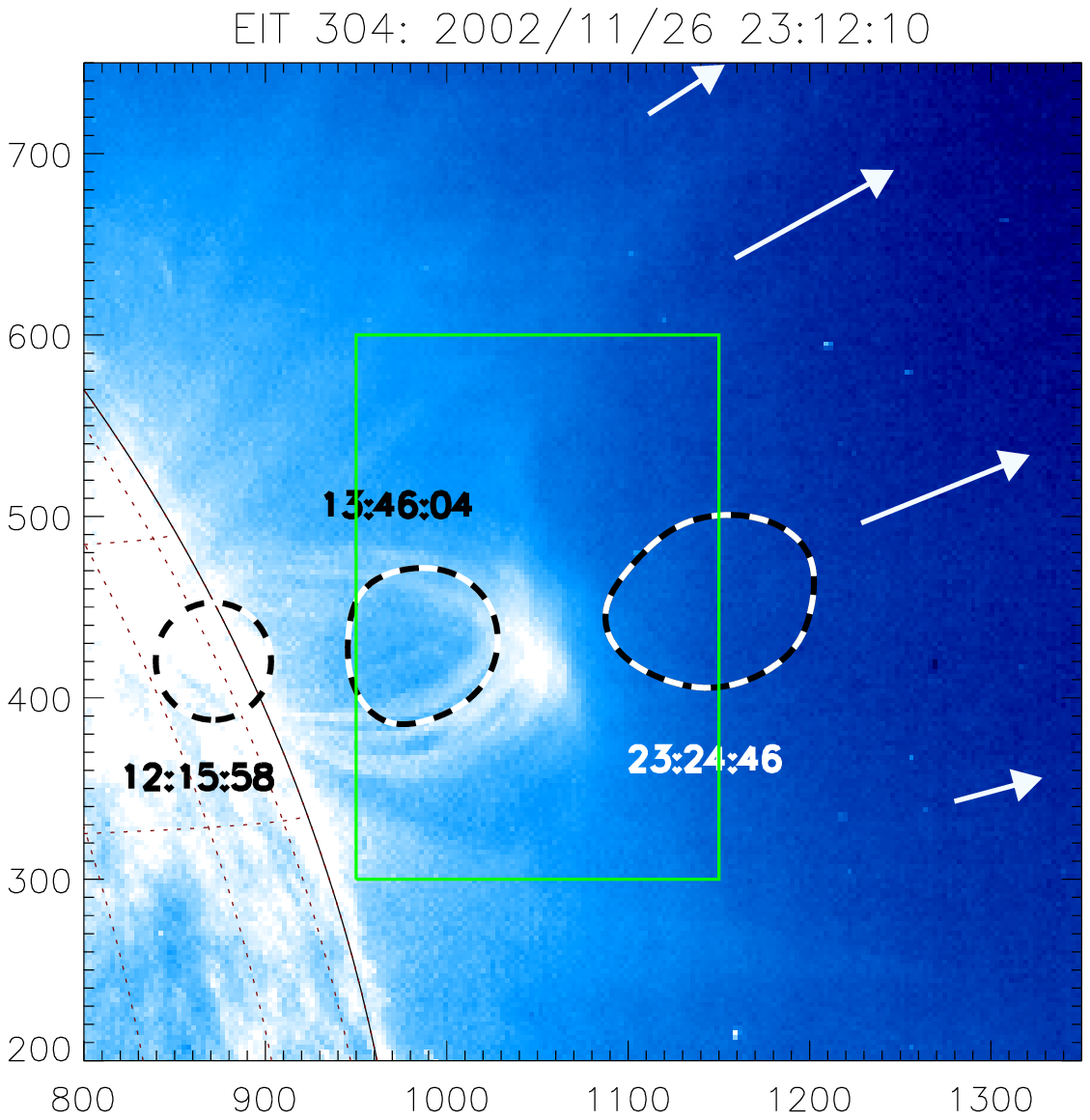}
		\caption{
			SOHO/EIT 304 {\AA} image taken at 23:12:10 on 2002 November 26, with RHESSI 6--12 keV contours (50\% level) at different times (5 minute exposures centered around 12:15:58, 13:46:04, \& 23:24:46 UT).
			Using values from Figure 6 and from Figure 9 (top left) of \citet{Bemporad2006},
			we have drawn arrows that delimit the angular extend of the region where UVCS observed hot plasma at 1.7 $R_S$.
			The {\it long} arrows delimit the region of plasma above 5 MK, while the {\it short} arrows delimit plasma above 3 MK.
			The green box is the ROI used to compute the {\it dark gray} lightcurve in the second plot of Figure~\ref{fig:tp}.
		}
		\label{fig:compositeimg}
		\end{figure}		

		At about 12:00 on November 26, a solar flare started (Figure~\ref{fig:tp}, first and third plots). 
		It was observed with RHESSI on the western limb of the Sun, at about 25 degrees of north latitude.
		RHESSI observed it until $\approx$12:42, at which time it entered Earth's shadow.
		Very shortly (3--4 minutes) after the rise in GOES fluxes at $\approx$13:40, RHESSI came out of Earth's shadow, and imaged an X-ray source
		at the same solar latitude, but about 80 Mm above the limb (Figure~\ref{fig:compositeimg}).
		This high altitude coronal X-ray source (hereafter HACXS) remains observable by RHESSI until $\approx$01:10 the next day (2002 November 27), i.e. for almost 12 hours.
		During that 12-hours period, several disc flares occurred, and with their much higher fluxes, often drowned the HACXS when attempting imaging.

		From its start at about 13:45, to about 16:00, the HACXS moved mostly radially outward from the Sun, at about 1.6 km/s (Figure~\ref{fig:tp}).
		During that time, the HACXS flux increased and then decreased.
		At about 16:15, the source seemingly ``jumps'' in altitude, by about 60 Mm.
		It could be argued that our initial source actually dimmed, and that this is a new, different source that appears at higher altitude.
		The RHESSI coverage between $\approx$16:00 and $\approx$17:00 is spotty: 	
		during that time interval, a flare occurred on the eastern limb of the Sun (introducing noise in images of our region-of-interest, ROI), 
		the spacecraft was initially in the South Atlantic Anomaly (with detectors turned off), and also spent time in Earth's shadow.
		The 5-minute image that shows a source midway between the two sites (at $\approx$16:20 in Figure~\ref{fig:tp})
		suggest we might indeed have had a single exciter that jumped across 60 Mm in about 70 minutes ($\approx$14 km/s velocity).

		Between about 16:00 and 18:45, the X-ray source moved progressively faster towards the solar equator (azimuthal velocity close to 5 km/s), 
		then stopped just as a flare at the footpoint of the loop system appears (position [900,400] in Figure~\ref{fig:compositeimg}).
		EIT images show the rise of a filament-like feature at around 16:12, from the same active region, 
		and a cusp-like feature and expelled material at around 17:12--17:24, the latter two in the same direction as the CME, starting below the HACXS altitude,
		and just a few tens of arcseconds northward of it\footnote{http://sprg.ssl.berkeley.edu/$\sim$shilaire/movies/20021126\_js/}.

		The X-ray source then settled on a mostly radial course at $\approx$2 km/s, before ceasing to be observed by RHESSI around 01:10 on 2002 November 27.
		This velocity is in very good agreement with the velocity of the rising post-CME loop systems observed with $EIT$ (Figure~\ref{fig:tp}).
		The HACXS stays well above ($\sim$0.1 $R_S$) the EUV loop system throughout the observations.
				
	\subsubsection{X-ray source size and shape:}

		\begin{figure*}[ht!]
		\centering
		\includegraphics[height=16cm]{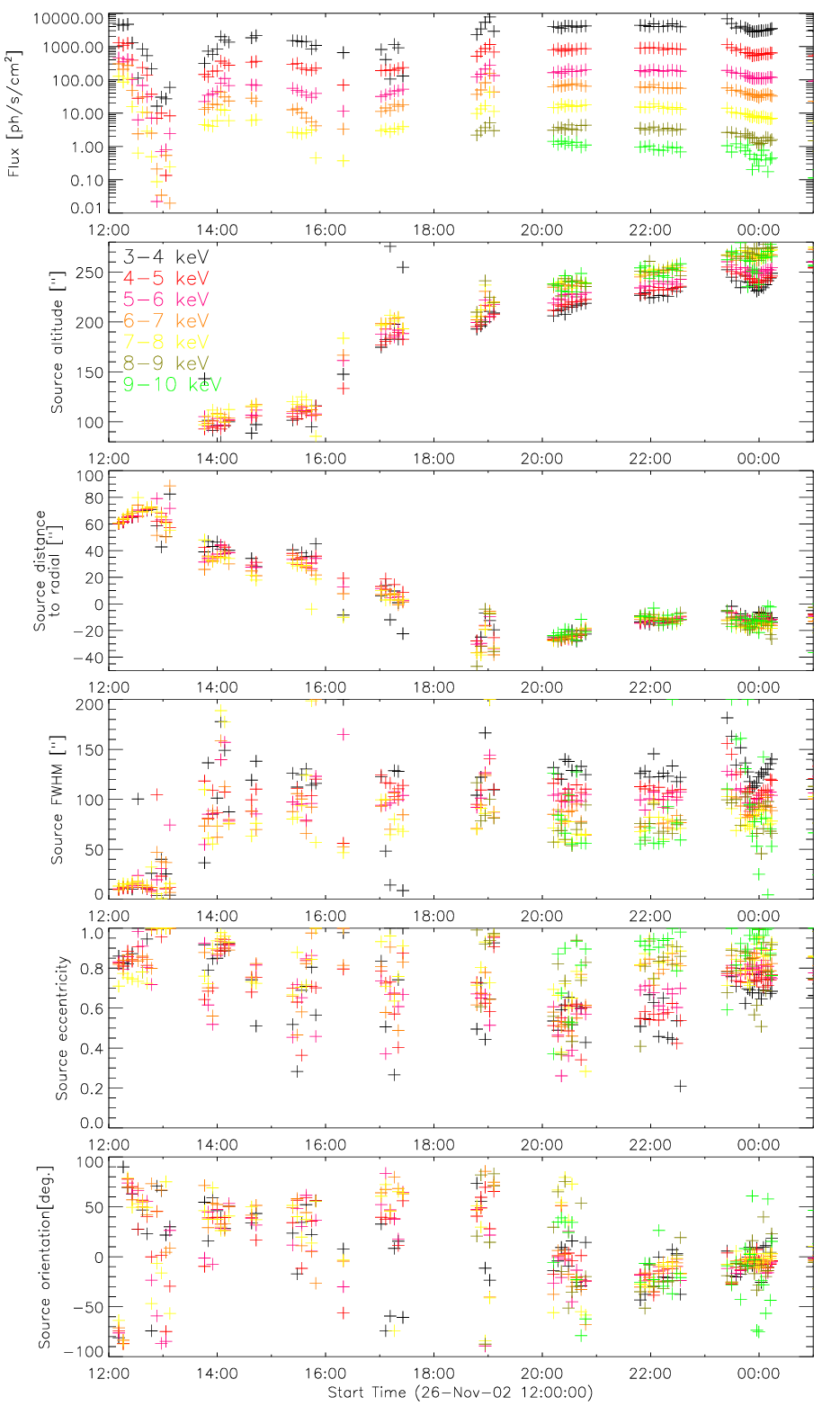}
		\caption{High altitude coronal X-ray source characteristics at different energies, derived from RHESSI visibilities accumulated over five minute intervals: 
			flux, altitude above photosphere, distance to $25^{\circ}$ radial; 2D Gaussian FWHM, eccentricity and orientation with respect to solar equator.
			For clarity, error bars were omitted, but the scatter of the points is a good approximation.
			Information on the 8--9 and 9--10 keV bands has been omitted before 18:00 and 20:00 UT, repectively, because of their weak fluxes.
		}
		\label{fig:visfittingtimeprof}
		\end{figure*}

		Using RHESSI visibilities \citep[a new software method akin to radio visibilities, see e.g.][]{Schmahl2003}, source size, shape, and position were determined and are displayed in Figure~\ref{fig:visfittingtimeprof} at different energies.
		Higher energies tend to be at higher altitude, suggesting that the hottest plasma is at higher altitude, as would be expected if the reconnection X-point already flew past our region of interest
		(see Section~\ref{sect:discussion} and Appendix~\ref{appendix:epslocation}).

		Source size does not vary remarkably during the 12-hours interval. The shape of the HACXS stays generally elongated, with the higher energies having the tendency for higher eccentricities.
		%It is to be noted that after a flare that occurred right at the footpoints of our loop system, around 18:30, the shape of the HACXS becomes more circular, before resuming its more linear aspect later on.
		%Moreover, the orientation of the source was a few degrees left of radial before the $\approx$18:30--19:15 period, and a few degrees right of radial afterwards.

	\subsubsection{Spectroscopy:} \label{sect:spectroscopy}

		\begin{figure*}[ht!]
		\centering
		\includegraphics[width=16.6cm]{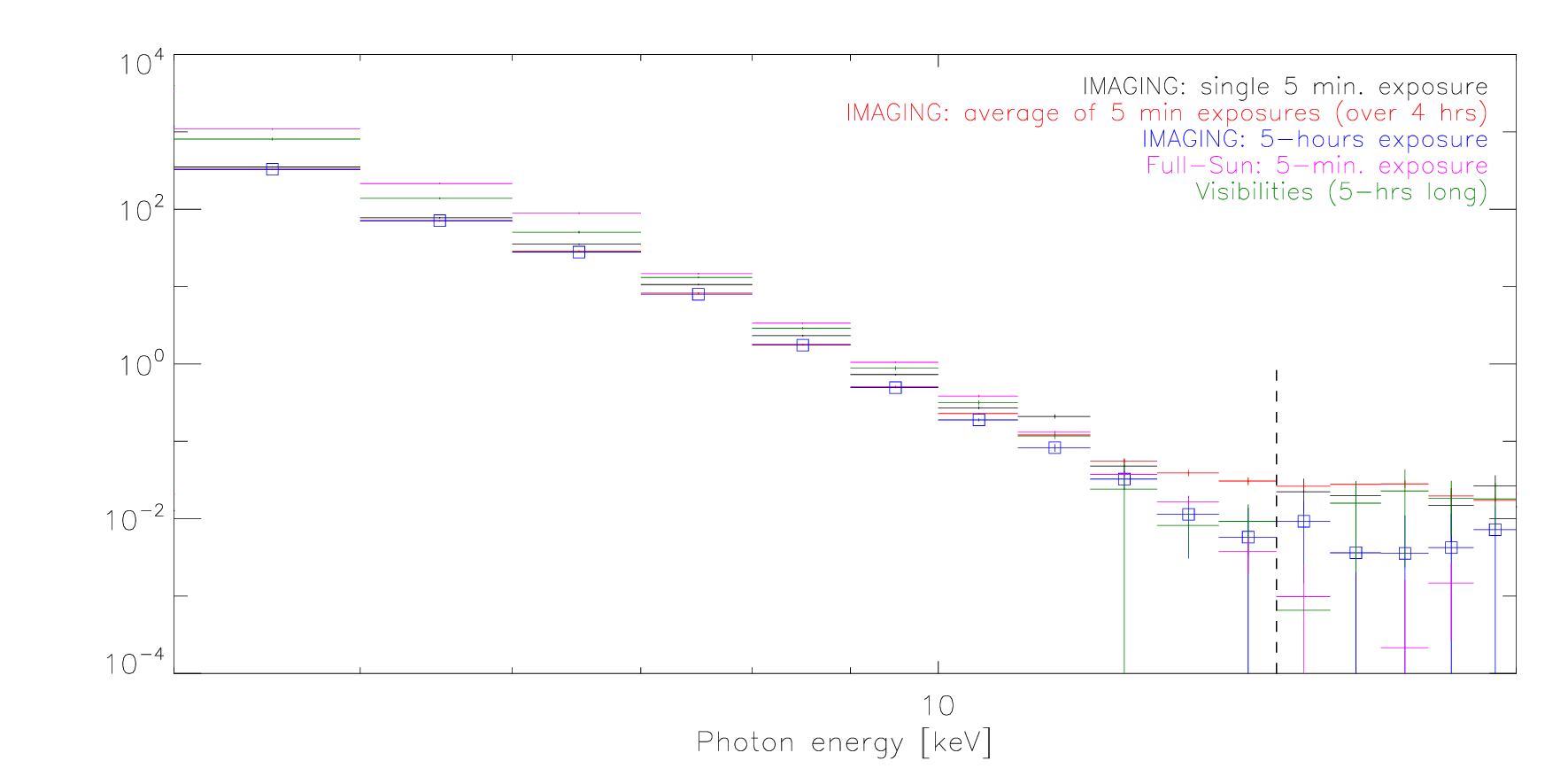}
		\caption{
			RHESSI spectra obtained using different methods.
			{\it Black data points}: RHESSI imaging spectroscopy with SC 8, accumulated between 2002/11/26 20:30:24 and 20:35:24 UT, with vertical error bars.
			The horizontal error bars actually correspond to the bin widths.
			{\it Red data points}: Average of all 5-minute imaging spectroscopy spectra (using SC 8) from 2002/11/26 20:00 to 24:00 UT.
			{\it Blue data points}: Imaging spectroscopy with SC 8, using the sum of all 5-minute images between 2002/11/26 20:00 and 24:00 UT. 
				Each image has been shifted in accordance with the source motion.
			{\it Purple data points:} Spatially-integrated spectroscopy between 2002/11/26 20:30:24 and 20:35:24 UT.
				Isothermal fitting yields 9.5 MK and 2.3$\times$10$^{47}$ cm$^{-3}$.
			{\it Green data points:} visibility-derived (shifted phase-centers) between 2002/11/26 20:10 and 2002/11/27 01:10 UT.
			Most methods yield unreliable values beyond $\approx$15 keV ({\it black vertical dashed line}), where background countrate is typically an order of magnitude above the source countrate.
		}
		\label{fig:spectroscopy}
		\end{figure*}

		\begin{figure}[ht!]
		\centering
		\includegraphics[width=8cm]{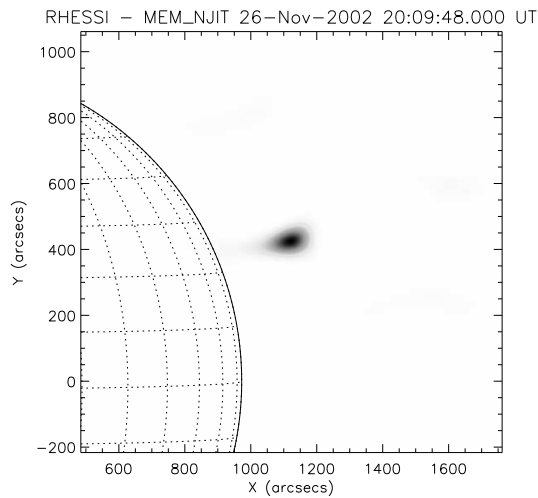}
		\caption{Five-hour long RHESSI image, from 2002/11/26 20:10 to 2002/11/27 01:10 UT, in the 4--8 keV energy band.
			RHESSI visibilities were generated in five minute accumulations, then bundled together after shifting their phases to remove source motion effects (relative to the source position at 20:09:48 UT).
		}
		\label{fig:longexp}
		\end{figure}

		Spectroscopy of our high-altitude X-ray source was done.
		In Figure~\ref{fig:spectroscopy}, full-Sun spectroscopy {\it (purple)} was done using the OSPEX {\it Solarsoft} suite of routines.
		Background selection and subtraction is a delicate process, particularly for that event, as disc flares occurred during the 12-hours time interval that the HACXS was observed.
		Hence, imaging spectroscopy was also employed:
		It is less sensitive than full-Sun spectroscopy, but does provide the inherent ability of removing background effects.
		The best spectrum from imaging was obtained by making 5-minute long images over five hours (2002 November 26 20:10 to 2002 November 27 01:10 UT), adding them together (rebinning and shifting for source motion), 
		and determining fluxes at different energies (i.e. spectrum) using back-projected maps with subcollimator 8 \citep{Hurford2002}: the noise level is typically 4 times smaller than from a spectrum
		obtained by simply adding together the spectra from each 5-minute accumulations.
		As an additional check, visibility-based imaging spectroscopy was also employed: the visibilities were phase-shifted to remove smearing from the source motion over five hours.
		The results are displayed as green data points in Figure~\ref{fig:spectroscopy}, and a long-integration image of our event is shown in Figure~\ref{fig:longexp}.
		The spectra displays no clear non-thermal (power-law) emission at high energies.

\begin{figure*}[ht!]
		\centering
		\includegraphics[width=10cm]{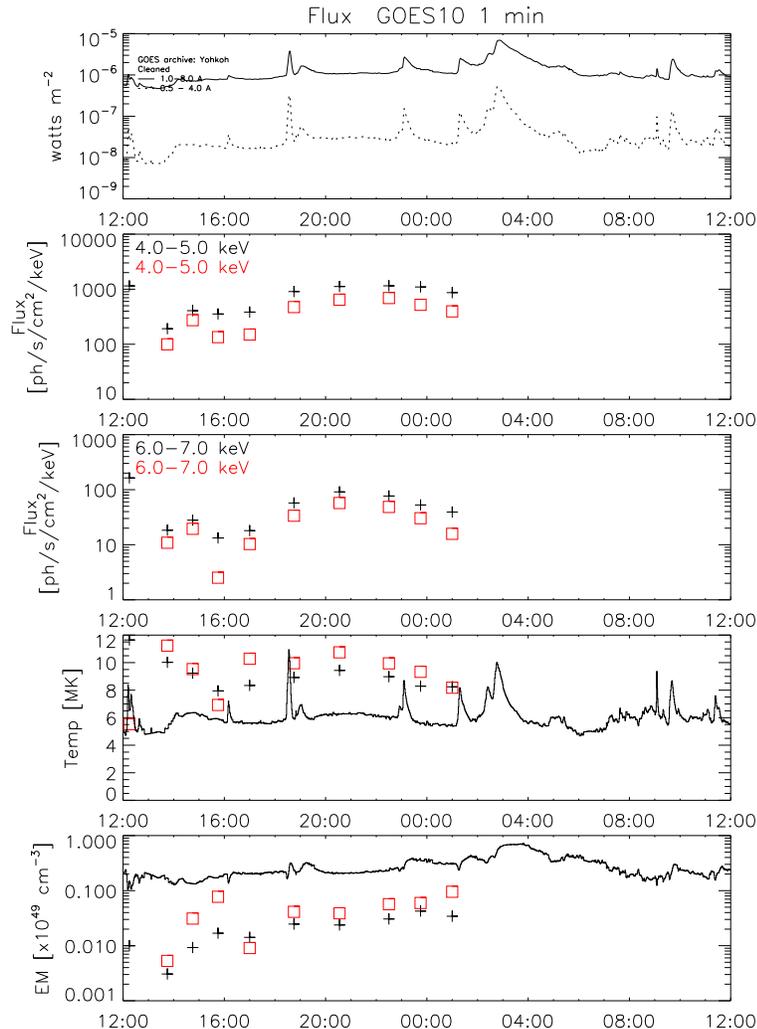}
		\caption{
			Black crosses: RHESSI full sun spectroscopy. Red squares: RHESSI imaging spectroscopy (SC 8).
			Solid black line: GOES temperature and emission measure measurements.
			The short duration ($\sim$minutes) peaks observed by GOES are due to disc flares, unrelated to the HACXS.
		}
		\label{fig:tem}
		\end{figure*}

		Fitting an isothermal component below 15 keV to the 5-hours back-projected data (blue data points in Figure~\ref{fig:spectroscopy}) yield 
		a temperature of $T$=11.4 MK and emission measure $EM$=1.4$\times$10$^{47}$ cm$^{-3}$.
		For comparison, 5-minutes long accumulation on 2002 November 26 around 20:30 UT yields $T$=11.5 MK and 1.2$\times$10$^{47}$ cm$^{-3}$ with imaging 
		and $T$=9.5 MK and 2.3$\times$10$^{47}$ cm$^{-3}$ with spatially-integrated spectroscopy (Figure~\ref{fig:spectroscopy}, {\it black} and {\it purple} data points).
		The lower temperature and higher emission measures obtained from RHESSI full-Sun spectroscopy are probably due to the presence of a low-energy flux component with large spatial extend, 
		the likely residuals from previous disc flares. 
		The even lower temperatures and higher emission measures measured by GOES ($T$=7.5 MK, $EM$=10$^{48}$ cm$^{-3}$) at the same time and throughout this event (Figure~\ref{fig:tem}) are typically attributed to the GOES response \citep[see e.g.][]{Holman2003}, 
		which is more sensitive to lower temperature plasmas.

		%CCL: Possible non-thermal source. $F_{50}$=7.5$\times$10$^{-6}$, $\gamma \approx$6 is an upper limit.
		%How does long-exposure imaging spectroscopy fare?

%+++++++++++++++++++++++++++++++++++++++++++++++++++++++++++++++++++++++++++++++++++++++++++++++++++++++++++++++++++++

%+++++++++++++++++++++++++++++++++++++++++++++++++++++++++++++++++++++++++++++++++++++++++++++++++++++++++++++++++++++
\section{Discussion} \label{sect:discussion}

\subsection{X-ray source timeline, position, and morphology}
	The HACXS first appeared some 80 Mm above the footpoints of a loop system, about 1.5 hours after a flare located in these footpoints erupted.
	It progressed generally outwards, its intensity rising and then decreasing over the course of $\approx$2 hours.
	From $\approx$17:00 to $\approx$19:30, EIT observed material being formed (e.g. a cusp feature and a filament feature) and expelled (e.g. a flux rope, and other ejecta).
	These ejections seem to have disturbed the HACXS: it jumped about 60 Mm in altitude, and its flux started increasing again.

	The height and velocity profiles of the HACXS and the EUV loop system support the picture of a looptop (or ``above the looptop'') reconnection point that moves upwards,
	heating the local plasma to X-ray emitting temperatures, before they cool down and are later seen in EUV, giving the impression that the EUV loops trail the X-ray source in space, 
	when in fact they are trailing in time.
	Apart from its very long duration, the source altitude profile of the HACXS is very similar to the observations reported by \citet{Gallagher2002}, including the higher energies being located at slightly higher altitudes than the lower energies.
	This further supports the scenario that hotter plasma is located at higher altitude (see Appendix~\ref{appendix:epslocation} for a simple justification, and Section~\ref{sect:epslocationmodeling} for an attempt at modeling it), 
	consistent with the \citet{Kopp1976} model and the \citet{Svetska1987} observations.

	After $\approx$21:00, the high-eccentricity (elongation) and orientation of the 2D Gaussians fitted to the HACXS (as shown in the two bottom plots of Figure~\ref{fig:visfittingtimeprof}) 
	are consistent with the geometry of a CS-like feature that extends radially outward.

	The HACXS ceased to be observed by RHESSI around 01:10 UT on 2002 November 27.
	This is due to both it having decreased in intensity to near or below RHESSI's sensitivity and the intense flaring activity that started at that time and lasted several hours.
	As observed in Figure~\ref{fig:goes}, the GOES baseline (i.e. non-flaring level) after 2002 November 27 $\approx$01:10 UT is ill-determined, because of the intense flaring activity.
	But it is conceivable that the HACXS remains present until 2002 November 29 $\approx$00:00, as the GOES X-ray flux in both GOES channels never drops back to pre-event levels until then.

\subsection{Energy-position relationship} \label{sect:epslocationmodeling}

	The peak emissions at different energies are slightly displaced (Fig.~\ref{fig:visfittingtimeprof}).
	In fact, \citet{Sui2003} and \citet{Liu2008} have observed similar behavior with RHESSI X-ray data:
	they observed the centroid position of successively higher energies to be located at higher altitude, and, then the trend reversed.
	They have attributed this behavior to a hot CS located at the position where the trend reversed. 
	Another possibility to explain the spatially-displaced energies is the presence accelerated particles which, much as in \citet{Brown2002}, 
	are being stopped at larger distances (column densities) the larger their initial energies are.
	Although the greater elongation observed at high energies supports that scenario \citep[see e.g. Appendix A of][]{PSH2009},
	the absence of non-thermal radiation (discussed in Section~\ref{sect:nth}) in the spatially-integrated spectrum clearly dispels that hypothesis.

\begin{figure*}[ht!]
		\centering
		\includegraphics[height=7cm]{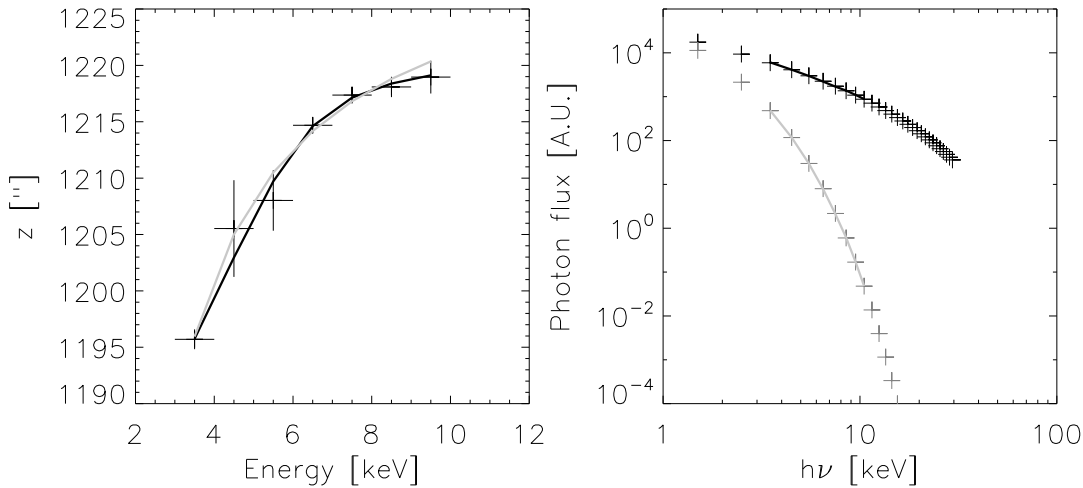}
		\caption{
			{\it \bf Left:}
				{\it Crosses:} Positions of emission at different energies, with error bars (vertical lines) and bin widths (horizontal lines).
			{\it Solid black line:} Fitting to the Gaussian $T$ and exponential $EM$ profiles.
			{\it Solid gray line:} Same, with $T$ fixed to 10 MK.
			{\it \bf Right:} 
			{\it Crosses:} Synthetic X-ray spectra generated using the fitting parameters found.
			{\it Solid line:} Fit to the crosses in the 4--10 keV band. Color scheme same as in the plot on the left.
			See text for more details.
		}
		\label{fig:tgrad2}
		\end{figure*}

	We have considered different models of temperature and emission measure profiles (see Appendix~\ref{appendix:epslocation}) to model the emission of 
	different energies at different positions.
	We have attempted to fit our RHESSI data (visibilities accumulated from 2002/11/26 20:10 to 2002/11/27 01:10, and phase-shifted to remove source motion smearing) 
	with the last two models mentioned in Appendix~\ref{appendix:epslocation}: the first one, using exponential profiles (with altitude) for both temperature and emission measure, yielded very poor results.
	The second one, where the temperature profile was assumed Gaussian, and the emission measure profile remained exponentially decreasing with altitude, yielded better results:
	The best-fitting parameters were $T_0$=119 MK, $H_T$=38.1'', $H_{EM}$=657'', $z_0$=1221.7'', and reduced $\chi^2$=0.33 (Figure~\ref{fig:tgrad2}, {\it black}).
	Such a high temperature seems highly unlikely.
	The synthetic X-ray spectrum computed from a plasma with such a temperature distribution (Figure~\ref{fig:tgrad2}, {\it} right) is clearly not observed.
	The height scales are loosely compatible with the typical density height scales found in the corona (about 100'').
	Given that our error bars are rather large (partly explaining the good chi-squared despite the obviously too-high temperature), 
	we have tried for comparison to fix the temperature $T$ at 10 MK, and redo the fitting process.
	We found $H_T$=4115'', $H_{EM}$=64525'', $z_0$=1232.8'', and reduced $\chi^2$=0.9 (Figure~\ref{fig:tgrad2}, {\it gray}).
	The corresponding synthetic spectrum is more in accordance with our observations, but these new rather large scales heights are somewhat unexpected,
	and would mean that the densities along the HACXS (and possibly the CS) change very slowly with altitude.
	
	A full exploration of the model space and other fitting techniques are beyond the scope of this paper, but will be addressed in a subsequent one.

\subsection{Energetics}

\subsubsection{Thermal energy in RHESSI source}  \label{Sect:th}
	We now want to estimate how much power and energy are needed to maintain the X-ray source at such high temperature for such a long time.
	On 2002 November 26, around 20:30, RHESSI observations indicate the source has an emission measure of about 2$\times$10$^{47}$ cm$^{-3}$,
	and a source size of $\approx$100'' FWHM, leading to a source volume $V$=2$\times$10$^{29}$ cm$^3$ (assuming HACXS has a spherical shape).
	Using $n=\sqrt{EM/V}$, one obtains an electron density $n$=10$^9$ cm$^{-3}$, and a total number of electrons $nV$=2$\times$10$^{38}$ electrons 
	(incidentally, a typical number for total flare-accelerated electrons found in HXR footpoints of large flares).
	With the temperature $T \approx$10.5 MK (average of the 9.5 MK and 11.5 MK found in Section~\ref{sect:spectroscopy}), this leads to a radiative loss timescale of $\tau_{rad}\approx$8$\times$10$^{4}$ s.
	The half loop-length from the chromospheric footpoint to the HACXS is $L\approx$$\pi/2 \times H$, with the source height $H$=220'', from which the conductive loss timescale $\tau_{cond} \approx$8$\times$10$^2$ s can be derived
	\citep[see e.g.][]{AschwandenBook2004}
	(we assumed energy is lost mainly to the chromospheric heat sink, and not to interplanetary space via open field lines).
	Conductive losses are more important than radiative losses, and we will hence use the former to estimate the power $P$ required to maintain the temperature of the high altitude X-ray source at around 10 MK, 
	using:
	\begin{eqnarray}
		P	&=&	\frac{E_{th}}{\tau_{cond}}	=	\frac{3kTnV}{\tau_{cond}}	\\
		 	&\approx&	6 \times 10^{35} keV/s  \approx	1 \times 10^{27} erg/s,
	\end{eqnarray}
	where $k$ is Boltzmann's constant, and $E_{th}$ is the thermal energy content of the source ($\approx$9$\times$10$^{29}$ erg around 20:30 UT).
	I.e. over 12 hours, about 4$\times$10$^{31}$ ergs must have been deposited in the source.

\subsubsection{Non-thermal energy} \label{sect:nth}

	If the power $P$ calculated in the previous paragraph came exclusively from accelerated particles as they dump all their energy into heating the plasma, 
	could it be that their associated non-thermal emission is so weak as to be unobservable by RHESSI?
	A photon power-law with spectral index $\gamma \approx$8 and flux at 10 keV $F_{10} \approx$3$\times$10$^{-2}$ photons s$^{-1}$ cm$^{-2}$ keV$^{-1}$ is an order of magnitude below the observed (thermal) X-ray emission (Figure~\ref{fig:spectroscopy}),
	and could be concealed by it. 
	The characteristics of the electron distribution corresponding to such a hypothetical non-thermal photon emission can be determined assuming either {\it thick-target} or {\it thin-target} assumptions:
	The column density traversed $N$=$n L$=2.5$\times$10$^{19}$ cm$^{-2}$ stops injected electrons with start energy below 11 keV.
	$N$= $n \sqrt[3]{V}$= 6.3$\times$10$^{18}$ cm$^{-2}$ stops injected electrons with start energy below 5.6 keV.
	I.e. we are very near a thick target at the energies where we have observed emission.
	Using Eq.~\ref{eq:nth:thick:2} of Appendix~\ref{appendix:brm} to obtain the low-energy cutoff value $E_C$ that equates thermal and non-thermal energies, we obtain $E_C$=6.3 keV.
	A plausible value, although flares have never been observed to go that low \citep[partly because thermal emission usually blocks any attempts at such observations, see e.g.][]{Holman2003,Kontar2008}.
	We find the total number of injected electrons in this case to be $F_{tot}=8.5 \times 10^{34}$ electrons per second above 6.3 keV. 
	This injection rate is at least an order of magnitude below typical large flare values \citep{Holman2003}.
	Assuming accelerated electrons escape (which need not be the case), this rate implies the HACXS must be replenished every $\approx$40 minutes,
	far from the ``flare number problem'', where the acceleration region is estimated to be replenished sometimes as fast as every few tens of seconds \citep[see e.g.][]{Miller1997}.

	A photon spectral index of $\gamma \approx$8 is not very flare-like, although not unlike what \citet{Liu2008} have found in coronal sources during the impulsive phase of a flare.
	Smaller $\gamma$ are permissible, but will decrease $E_C$ correspondingly, in order to conserve $P$ to the same required amount.
	$E_C$ cannot be below $\approx$1 keV, the thermal temperature of the plasma: at these energies, electrons would essentially be indistinguishable from the local thermal plasma, and not contribute energy to a non-thermal beam of electrons \citep{Emslie2003}.
	For $E_C$=1 keV, a $\gamma$=3.2 non-thermal power-law would conserve the injected non-thermal power and still be concealed below the thermal emission.

	While it is to be noted that low-energy cutoffs below 10 keV have so far never been reliably observed,
	the heating of the HACXS over 12 hours purely by accelerated electrons cannot be firmly contradicted by our observations.	

\subsection{Connection between RHESSI observations and UVCS observations}

	There is a small overlap in time when both RHESSI observes the bottom of the CS (at an altitude of about 0.3 $R_S$), 
	and UVCS can make a reliable temperature diagnostic of the CS at 0.7 $R_S$:
	On 2002 November 27, between 00:00 and 01:10,  both RHESSI and UVCS sources indicate temperatures of $\approx$8 MK.
	%RHESSI indicates $EM$=5$\times$10$^{47}$ cm$^{-3}$.
	%UVCS indicates: $EM$=$n^2V$=7$\times$10$^7$ cm$^{-3}$ $\times$ 10$^4$ km $\times$ 28'' $\times$ 100 Mm $\approx$ 10$^{44}$ cm$^{-3}$.
	If one infers that this temperature is constant between these two altitudes, one can estimate the total amount of (thermal) energy contained within this region.
	
	Assuming a rectangular sheet with $\approx$100" width, $\approx$0.4 $R_S$ length, $\approx$10$^4$ km thickness \citep[as assumed by][]{Bemporad2006},
	then one gets a volume $\approx$2$\times$10$^{29}$ cm$^3$.
	Assuming the average electron density to be the geometric mean between what is found by RHESSI at 0.3 $R_S$ ($\approx$10$^9$ cm$^{-3}$) and what is found by UVCS at 0.7 $R_S$ \citep[$\approx$ 7$\times$10$^7$ cm$^{-3}$, according to][]{Bemporad2006},
	one finds a total of $\approx$6$\times$10$^{37}$ electrons, for a total thermal energy content of $\approx$4$\times$10$^{37}$ keV, or $\approx$6$\times$10$^{28}$ erg,
	i.e about 7\% of the instantaneous thermal energy found in the HACXS (and about 0.1\% of the total energy that must have been injected in the HACXS over 12 hours).

	Hence, it is conceivable that the energy that powers the CS comes from the HACXS region,
	e.g. via heat conduction, and not only via magnetic reconnection in the CS.
	On the other hand, the fact that the HACXS starts before the CME ($\approx$13:40 vs. $\approx$17:00), 
	the fact that \citet{Bemporad2008}'s non-thermal turbulent reconnection model explains well the UVCS observations,
	and the fact that the expelled material observed in EUV appears to flow beside the EUV loop system and the HACXS,
	still leaves the question open as to whether the CME/CS and the HACXS are significantly tied together.

%+++++++++++++++++++++++++++++++++++++++++++++++++++++++++++++++++++++++++++++++++++++++++++++++++++++++++++++++++++++
\section{Summary and Conclusion}

	UVCS observations in the wake of a CME that started around 2002 November 26 17:00 UT show hot plasma (initially well over 8 MK) at 0.7$R_S$ above the photosphere, 
	for 2.3 days (at which point it had cooled down below $\approx$ 3.5 MK and was no longer observed).
	This hot plasma was interpreted as the signature of current sheet material.

	X-ray observations during the same time interval show enhanced X-ray emissivity throughout that period, albeit at lower altitudes ($\lesssim$0.3 $R_S$).
	For 12 hours, RHESSI observes a thermal coronal source that is near the base of the curent sheet,
	and, as it has at least an order of magnitude more thermal energy, it could technically provide the heat for the CS
	(i.e. in this scenario, reconnection and plasma heating occur mostly near the looptops rather that in the CS).
	On the other hand, heat and energy transport through a turbulent environment (as is probably the CS) can be quite complex and slow 
	(usual plasma coefficients must be replaced by effective ones: anomalous heat conductivity), 
	and as turbulent reconnection in the CS \citep{Bemporad2008} explains elegantly the CS temperature, 
	it is possible that the coronal X-ray source and EIT looptop system are only weakly related to the CS/CME.
	
	We used novel long-accumulation imaging spectroscopy techniques to better estimate the photon spectrum and we have fitted it with an isothermal component.
	The RHESSI source temperature peaked at 10--11 MK.
	The emission measure of this source essentially increased during this whole 12-hours period, reaching above 5$\times$10$^{47}$ cm$^{-3}$.

	We have also observed an energy vs. position displacement in the emission from this HACXS, 
	consistent with a plasma that has a Gaussian profile for its temperature distribution with altitude.

	Because of the lack of observed non-thermal emission, it appears unlikely, though not impossible, 
	that the heating in the HACXS is due to particles being accelerated in it.

	%We found that it was possible for the heating of the high altitude coronal X-ray source to be the result of particle acceleration in the source, 
	%and for their non-thermal bremsstrahlung spectrum to be unobserved.
	%This heating could in turn be transferred to the current sheet plasma trailing the CME,
	%and is an alternative to \citet{Bemporad2008}'s explanation for the high current sheet temperatures.

%==================================================================================================================	
%==================================================================================================================	
%==================================================================================================================	
\appendix
%==================================================================================================================	
\section{Thermal bremsstrahlung and emission energy gradient}\label{appendix:epslocation}

	The photon flux of an isothermal plasma of temperature $T$ and emission measure $EM$ is given by:
	\begin{equation}
		I(\varepsilon) = C_{thermal} \, EM \, \frac{1}{\varepsilon} \, \frac{e^{-\varepsilon/kT}}{\sqrt{kT}}	
	\end{equation}
	where $k$ is Boltzmann's constant, $\varepsilon$ is the observed photon energy, and the constant, $C_{thermal}$=1.54$\times$10$^{-42}$ photons s$^{-1}$ cm keV$^{1/2}$.
	In the following, both the emission measure $EM$ and the temperature $T$ will be assumed to be functions of $z$, the altitude above the photosphere.
	Hence, $I(z,\varepsilon)$ reaches a maximum along $z$ when $\frac{dI}{dz}=0$. 
	\begin{eqnarray}
		\frac{dI}{dz}(z,\varepsilon)	&=&	\frac{\partial EM}{\partial z}	\frac{1}{\varepsilon} \frac{e^{-\varepsilon/kT}}{\sqrt{kT}} + \frac{EM}{\varepsilon} \frac{\partial}{\partial z} \left( \frac{e^{-\varepsilon/kT}}{\sqrt{kT}} \right) \\
						&=&	\frac{\partial EM}{\partial z}	\frac{1}{\varepsilon} \frac{e^{-\varepsilon/kT}}{\sqrt{kT}} + \frac{EM}{\varepsilon} \frac{\partial kT}{\partial z} \frac{\partial}{\partial kT} \left( \frac{e^{-\varepsilon/kT}}{\sqrt{kT}} \right) \\
						&=&	\frac{\partial EM}{\partial z}	\frac{1}{\varepsilon} \frac{e^{-\varepsilon/kT}}{\sqrt{kT}} + \frac{EM}{\varepsilon} \frac{\partial kT}{\partial z} \frac{e^{-\varepsilon/kT}}{(kT)^{3/2}} \left( \frac{\varepsilon}{kT} - \frac{1}{2} \right) \label{eq:dI_dz:1} \\
						&=&	\frac{1}{\varepsilon} \frac{e^{-\varepsilon/kT}}{\sqrt{kT}} \left[ \frac{\partial EM}{\partial z} + \frac{EM}{kT} \frac{\partial kT}{\partial z} \left( \frac{\varepsilon}{kT} - \frac{1}{2} \right)	\right]	
	\end{eqnarray}
	I.e. $\frac{dI}{dz}$=0 when:
	\begin{equation}
		\frac{1}{EM} \frac{\partial EM}{\partial z} + \frac{1}{T} \frac{\partial T}{\partial z} \left( \frac{\varepsilon}{kT} - \frac{1}{2} \right) = 0
		\label{eq:dI_dz:2}
	\end{equation}

	Let us briefly study a few solutions that satisfy Eq.~(\ref{eq:dI_dz:2}).

	\paragraph{Assuming $T=const$ along $z$:} 
	Then $\frac{dI}{dz}=0$ when $\frac{\partial EM}{\partial z} = 0$, i.e. $EM$ is also constant along $z$. 
	Emission at all energies is hence constant along $z$.
	%This solution is of little physical interest.

	\paragraph{Assuming $EM=const$ along $z$:} 
	Then $\frac{dI}{dz}=0$ when $\frac{1}{T} \frac{\partial T}{\partial z} \left( \frac{\varepsilon}{kT} - \frac{1}{2} \right)= 0$.
	If $T(z)$ is monotonic, then this reduces to $\left( \frac{\varepsilon}{kT} - \frac{1}{2} \right)$=0, which leads to the conclusion that 
	not only are the higher energies emitted from regions of higher temperatures, but that these temperatures are of the order of $\varepsilon$.
	A non-monotonic $T(z)$ can modify that behavior somewhat.

	\paragraph{Assuming exponential profiles for $T$ and $EM$:}
		\begin{eqnarray}
			 T(z)	&=&	 T_0 \, e^{(z-z_0)/H_T}	\\
			EM(z)	&=&	EM_0 \, e^{-(z-z_0)/H_{EM}}	
		\end{eqnarray}
	where $T(z)$ and $EM(z)$ are the temperatures and differential emission measures along (a loop) path $z$.
	$T_0$ and $EM_0$ are constants. $H_T$ and $H_{EM}$ are scale heights.
	The peak of emissions at energy $\varepsilon$, located at $z$ are expected to follow the functional relation:
		\begin{equation}
			z-z_0	=	H_T \ln \left( \frac{\varepsilon}{kT_0 \, (1/2+H_T/H_{EM})} \right)
		\end{equation}
	These profiles diverge, and are of course globally unphysical. At best, this model can only apply locally.
	
	\paragraph{Assuming Gaussian profile for $T$ and exponential for $EM$:}
		\begin{eqnarray}
			 T(z)	&=&	 T_0 \, e^{-\left((z-z_0)/H_T\right)^2}	\\
			EM(z)	&=&	EM_0 \, e^{-(z-z_0)/H_{EM}}	
		\end{eqnarray}
	which leads to a slightly more complicated functional relationship:
		\begin{equation}
			\varepsilon = \frac{kT_0}{2} e^{-\left( \frac{z-z_0}{H_T} \right)^2} \left( 1-\frac{H_T^2}{H_{EM}} \frac{1}{z-z_0} \right) 
		\end{equation}

%==================================================================================================================	
\section{Non-thermal emission formulae}\label{appendix:brm}

\subsection{Thick-target assumption} 
	\citep{Brown1971,Hudson1972}
	The non-thermal photon power-law produced by the power-law distribution of electrons that produced it can be related by the following formula:
	\begin{eqnarray}
		\Phi_{thick}(\varepsilon) &=&	C_{thick}(\delta) \frac{F_{tot}}{(\delta-2)} E_c^{\delta-1} \varepsilon^{1-\delta} \\
					&=&	C_{thick}(\delta) \frac{P_{nth}}{(\delta-1)} E_c^{\delta-2} \varepsilon^{1-\delta} \label{eq:nth:thick:1}
	\end{eqnarray}
	where $\Phi_{thick}(\varepsilon)$ is the photon flux at 1 AU, 
	for photon energy $\varepsilon$ (in keV),
	in photons s$^{-1}$ cm$^{-2}$ keV$^{-1}$.
	$C_{thick}(\delta)=1.5 \times 10^{-34} B(\delta-2,1/2)$, 
	$B$ is the Beta function,
	$F_{tot}$ is the total number of electrons per second above the cutoff energy $E_c$,
	$\delta$ is the injected electron spectral index, and is equal to $\gamma+1$,
	where $\gamma$ is the photon power-law spectral index. 
	$P_{nth}=\frac{\delta-1}{\delta-2} \, E_c \, F_{tot}$ the non-thermal power in accelerated electrons, in keV/s.

	Eq.~(\ref{eq:nth:thick:1}) can also be rewritten:
	\begin{equation}
		E_c = \left( \frac{6.7\times10^{33}}{B(\gamma-1,1/2)} \, \gamma \, \varepsilon^{\gamma} \, \frac{\Phi(\varepsilon)}{P_{nth}} \right)^{\frac{1}{\gamma-1}}
		\label{eq:nth:thick:2}
	\end{equation}

\subsection{Thin-target assumption} 
	Using the same notations as in the preceding paragraph:
	\begin{eqnarray}
		\Phi_{thin}(\varepsilon)	&=&	C_{thin}(\delta) N (\delta-1) F_{tot} E_c^{\delta-1} \varepsilon^{-1-\delta}	\\
						&=&	C_{thin}(\delta) N P_{nth} E_c^{\delta-2} \varepsilon^{-(\delta+1)}	\label{eq:nth:thin:1}
	\end{eqnarray}
	$N$ is the column density traversed by the electrons (cm$^{-2}$),
	$C_{thin}(\delta)$=4.05$\times$10$^{-52} \frac{B(\delta,1/2)}{\delta}$,
	$\delta=\gamma-1$.
	This formula is valid when $\varepsilon_{keV} \gg \left( \frac{N}{2\times10^{17} cm^{-2}} \right)^{1/2}$.

	Eq.~(\ref{eq:nth:thin:1}) can be rewritten:
	\begin{equation}
		E_c= \left( \frac{2.5 \times 10^{51}}{B(\gamma-1,1/2)} \frac{\gamma-1}{\gamma-3} \varepsilon^{\gamma} \frac{\Phi(\varepsilon)}{N \, P_{nth}} \right)^{\frac{1}{\gamma-3}}
		\label{eq:nth:thin:2}
	\end{equation}

%==================================================================================================================	
\bibliographystyle{apj}
\bibliography{psh_biblio}

%+++++++++++++++++++++++++++++++++++++++++++++++++++++++++++++++++++++++++++++++++++++++++++++++++++++++++++++++++++++

%% Included in this acknowledgments section are examples of the
%% AASTeX hypertext markup commands. Use \url without the optional [HREF]
%% argument when you want to print the url directly in the text. Otherwise,
%% use either \url or \anchor, with the HREF as the first argument and the
%% text to be printed in the second.

\acknowledgments

This work was supported by NASA Heliospheric Guest Investigator grant NN07AH74G.
We would like to thank the anonymous referee for his or her time and extremely useful remarks, 
which greatly improved this work.


%% in parentheses, after the keyword, but they will not be verified.

%{\it Facilities:} \facility{RHESSI}, \facility{Hinode (XRT)}, \facility{GOES}, \facility{FOXSI}.



\end{document}